\documentclass[10pt,times,twocolumn]{article}

\usepackage[all]{xy}
\usepackage{color,url}
\usepackage{subfigure}
\usepackage{bm,amsmath}
\usepackage{marvosym}
\usepackage{graphicx}
\usepackage{wrapfig}
\usepackage{epsf}
\usepackage{times,mathptm}
\usepackage{url}

\def\(({\left(}
\def\)){\right)}
\def\[[{\left[}
\def\]]{\right]}

\newcommand{\be}{\begin{equation}}
\newcommand{\ee}{\end{equation}}
\newcommand{\bea}{\begin{eqnarray}}
\newcommand{\eea}{\end{eqnarray}}

\begin{document}

\title{\Large \bf Message Passing for Integrating and Assessing Renewable Generation\\ in a Redundant
Power Grid}

\author{{\bf Lenka Zdeborov\'a}, $^{(1)}$ {\bf Scott Backhaus} $^{(2)}$ and {\bf Michael Chertkov} $^{(1)}$\\
Center for Nonlinear Studies and Theoretical Division $^{(1)}$ and \\
Material Physics and Applications Division, $^{(2)}$ LANL, Los Alamos, NM 87545}

\maketitle

\begin{center}
{\bf \large Abstract}
\end{center}
{\it A simplified model of a redundant power grid is used to study
integration of fluctuating renewable generation.  The grid consists of large number of
generator and consumer nodes.  The net power consumption is determined by the difference between the gross consumption and the level of renewable generation.  The gross consumption is drawn from a narrow distribution representing the predictability of aggregated loads, and we consider two different distributions representing wind and solar resources. Each generator is connected to $D$ consumers, and redundancy is built in by connecting $R\leq D$ of these consumers to other generators.  The lines are switchable so that at any instance each consumer is connected to a single generator. We explore the capacity of the renewable generation by determining the level of  "firm" generation capacity that can be displaced for different levels of redundancy $R$. We also develop message-passing control algorithm for finding switch settings where no generator is overloaded.}


\vspace{0.5cm}

\section{Introduction}\vspace{-0.1cm}
The design and control of the existing power grid is mostly based upon centralized power
generation and one-way flow of power.  The model works quite well because inter-hour
loads variations are relatively predictable allowing reliable centralized scheduling,
and intra-hour load fluctuations are typically small and can be followed efficiently by rapid response
regulation reserves.  However, the effects of time-variable renewable generation are
beginning to impact this model, and the anticipated growth of wind and photo-voltaic (PV)
generation at the utility and residential scale could severely impact the reliability of
electrical power systems \cite{08EERE}.

Wind and PV generation are viewed by most utilities and electrical transmission
operators as "energy" sources, i.e. the electrical system will
accept the energy that these sources produce when the wind is blowing or the sun is
shining, but fluctuations in the output of these sources are large so that they are not
relied upon.  This is in contrast to a "capacity" source such as fossil, nuclear, or
hydroelectric generators whose output can be controlled and scheduled to follow the
inter-hour load variations.  When wind and PV are viewed strictly as an energy sources, no additional backup generation capacity needs to be added.  However, the addition of intermittent renewable generation affects the mix and dispatch of other generating and regulating reserves, increasing the grid's operating cost and the reserve margin \cite{Soder93,Dany01,DM05,07SPA}.  In essence, the apparent capital and operating cost of renewable generation is increased.

For wind and PV to make up a significant component of the generation mix, advances in
our understanding, planning, and design must be made so that some fraction of
this new generation can also be counted upon as a capacity source \cite{WDT84,AR91,BG08}.  Decreasing the variability, or dispersion, of these resources through geographical \cite{08GBK} and resource-base diversification or short-term storage \cite{NREL08PVST,KCR99} are possible routes receiving attention.

We explore an alternative method that utilizes switchable redundant transmission or distribution lines allowing for reconfiguration of the electrical network to match available generation supply to demand without resorting to load shedding  of any type, e.g. line tripping, demand response, etc.  Specifically, our prime focus is on preventing overloads of the firm power generation units caused by fluctuations in the demand and intermittent renewable generation by effectively utilizing ancillary lines.  Although the current transmission grid is high meshed, we consider a modification where certain lines are switchable so that the operational realization of the transmission grid is kept radial or tree-like to avoid power circulation and certain congestion issues associated with meshed grids.  Our simplified model is perhaps closer to many distribution systems where switchable redundant lines of this type already exist and could be expanded to imitate the type of system we suggest. Note also  that, as argued in \cite{08HOFO}, switching can play an essential role in relieving congestion on both transmission and distribution levels.

We explore this method by incrementally adding renewable generation to our simplified electrical model and determining the
minimum required firm generation to ensure that all loads can be satisfied.  The
reduction in the required firm generation as a fraction of the added renewable
generation provides a measure of its capacity.  We determine the renewable generation
capacity for different levels of redundancy as well as for different amounts of
generation variability.

The simplifications we make to create a tractable model are extreme.  We completely ignore
(a) power losses, line impedance, reactive power flow, and transient effects
\cite{96WW}, essentially modeling electricity delivery as an abstract commodity flow
\cite{90CLR}, (b) long distance exchange of active power that occurs in a meshed
transmission-level grid, (c) inhomogeneities and spatio-temporal correlations in loads
and generation, and (d) all economic, pricing and regulation effects \cite{01BH}. In this very simplified model, we strive to establish the fundamental limits or bounds on the capacity of renewable generation and renewable penetration levels beyond which no further capacity value is added.

The material in the paper is organized as follows.
In Section \ref{sec:Toy}, we introduce our simplified model of the electrical grid and
describe the set of problems addressed in this paper. In Section \ref{sec:results}, we
discuss simple bounds on the level of firm generation required for avoiding load shedding. Accurate limits of where redundancy allows the avoidance of load shedding within our model are established in Section \ref{sec:BP}. We assume that the graph is sparse and apply methods of Belief Propagation (BP) to create an efficient message-passing algorithm to find an acceptable grid configuration among the huge number of possibilities. This algorithm, which can also be used to control the grid,  is discussed in Section \ref{sec:Control}. Section \ref{sec:Con} summarizes our approach and proposes future extensions of the study.

\section{Grid model}
\label{sec:Toy}

Consider a crude simplification of the power grid with $M$ sources/generators each connected to $D$ distinct consumers, so that the total number of consumers is $N=M D$. Greek/Latin indices will be reserved for generators/consumers. For simplicity we assume that each generator has a maximum production rate of $\hat{y}$, i.e. ${\bm y}=(0\leq y_\alpha\leq \hat{y}|\alpha=1,\cdots,M)$, though inhomogeneities in the production can be easily incorporated into our approach.
The configuration of loads, ${\bm x}=(x_i>0|i=1,\cdots,N)$, is drawn from an assumed known distribution.
As an example of the load distribution we will consider the joint probability distribution of demand, where each load comes from a uniform distribution, centered around $1$ with width $\Delta$:
\begin{eqnarray}
\!{\cal P}({\bm x})\!=\!\prod_i p(x_i),\ \ p(x)=\left\{
\begin{array}{cc}
1/\Delta, &\!\!\! |x-1|<\Delta/2\\
0, &\!\!\! \mbox{otherwise}
\end{array}.
\right. \label{Px}
\end{eqnarray}
We generally assume that $\Delta \ll 1$ reflecting the relative predictability of electrical loads.

We also assume that  consumers  have the capability to generate electrical power. We consider two different renewable generation models; one spatially homogeneous and the other inhomogeneous. In the homogeneous model, renewable generation with an installed (nameplate) capacity $\hat{z}$ is present and operational at every node but the output of that generator can fluctuate between $0$ and $\hat{z}$ with equal probability, i.e. ${\bm z}=(0\leq z_i\leq \hat{z}|i=1,\cdots,N)$ is selected from the distribution
\begin{eqnarray}
{\cal P}({\bm z})=\prod_i p(z_i),\quad p(z)=\left\{
\begin{array}{cc}
1/\hat{z}, & 0\leq z \leq \hat{z}\\
0, & \mbox{otherwise}.
\end{array}
\right. \label{Pz}
\end{eqnarray}
The large width of ${\cal P}({\bm z})$ compared to ${\cal P}({\bm x})$ reflects the intermittent and highly variable nature of any one renewable generation installation in comparison to the relatively predictable loads.  The homogeneous model is representative of wind farms scattered over a relatively large area with the distribution ${\cal P}({\bm z})$ describing the variability of the wind resource between the different wind farms.  The form of ${\cal P}({\bm z})$ in Eq.~\ref{Pz} is just an example for this article.  Our methods are amenable to any bounded distribution.

In the inhomogeneous model, we consider that the output of the renewable generation at each node is either zero or equal to the nameplate capacity, i.e.
\begin{equation}
{\cal P}({\bm z})=\prod_i p(z_i),\quad p(z_i)=f_i\delta(z_i-\hat{z})+(1-f_i)\delta(z_i),\label{Pz2}
\end{equation}
where $f_i=0$ or $1$.  An example of the inhomogeneous model is  similar  to the homogeneous model where we replace the wind farms with PV farms and the $f_i$ indicates whether sun at farm $i$ is obscured by clouds or not.  In our modeling the $f_i$ are assigned randomly throughout the grid, and we represent the aggregate effect through a single variable $f=\Sigma f_i/N$ indicating the fraction of renewable generation  in operation.

We model the  transmission of electrical power from generators to consumers as an abstract commodity flow ignoring all effects of AC power flow, any correlations between fluctuations in load or renewable generation at different consumers, and any economic, pricing and regulation effects \cite{01BH}.  We start with a simple separated grid where a group of $D$ consumers is connected to their "local" generator, there being $M$ such groups.

\begin{figure}[ht]
\subfigure[\hspace{0.3cm} $R=0,1,2,3$. Graph samples. Ancillary connections to foreign generators/consumers are shown in color.]
{ \includegraphics[width=0.9\linewidth,page=1]{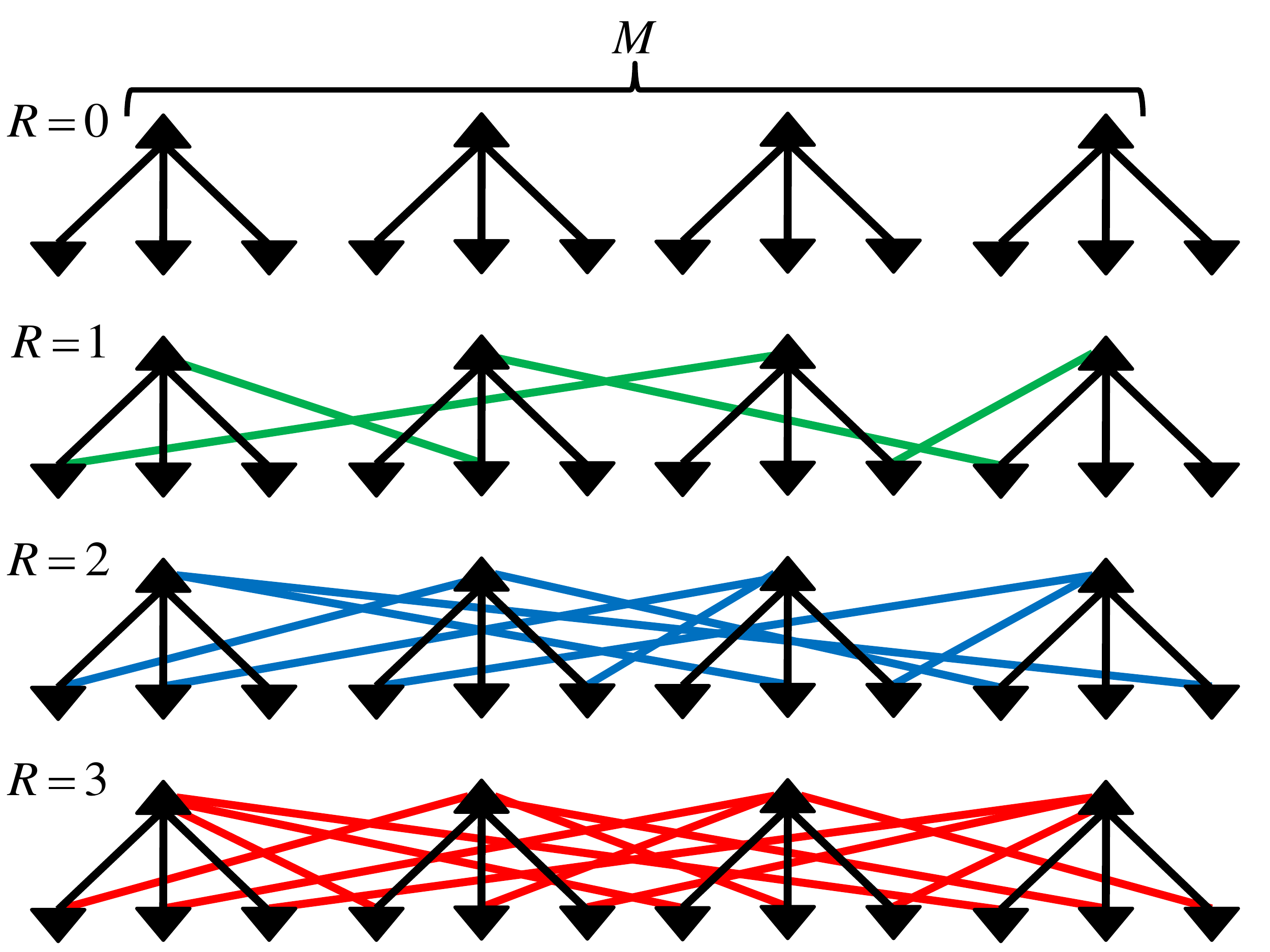}}
\subfigure[\hspace{0.3cm} $R=1$. Three valid (SAT) configurations (shown in black, the rest is in gray) for a sample graph shown in Fig.~\ref{graphs}a. ]
{\includegraphics[width=0.9\linewidth,page=2]{graphs.pdf}}
\caption{   Illustration of the bipartite graph construction of our grid with $D=3$.
\label{graphs}}
\end{figure}

To explore how redundancy affects the capacity value of the renewable generation, we consider an interconnected grid built from the separated one by adding ancillary lines between consumers and generators. For our quantitative analysis we choose a random graph drawn uniformly at random from the following ensemble.  Of the $D$ consumers connected to each local generator, $D-R$ of them are still simply connected to the local generator while $R$ of them are also connected to a second "remote" generator via an ancillary line, in such a way that each firm generator is connected to exactly $D+R$ consumers. In total, $MR$ consumers are connected to two generators and $M(D-R)$ are connected to one generator. See Fig.~\ref{graphs} for an illustration. Although some consumers are connected to two generators, the connections are assumed to be switchable so that only one is utilized so that the resulting "operational" grid is radial or tree-like to avoid power circulation and certain congestion issues associated with meshed grids.

Without being able to resort to load shedding, we consider the possibility of redistributing the load via the system of ancillary lines not used under normal (no overload) conditions. Fundamentally,  we ask the following question: can an intelligent
arrangement of ancillary lines among a system of $M$ generators  possibly decrease the amount of  firm generation required system wide by giving some consumers the ability to choose a connection to a generator  other than its local generator.  We judge the benefit of a certain number of redundant ancillary lines by computing the minimum firm generation capacity required to drive the probability of a firm generator overload from a finite number
to zero in the limit of $M\to\infty$  and $D=O(1)$.  The notion that a
drastic reduction in firm generation is possible stems from basic information-theoretic intuition: any finite error probability can be reduced to zero via properly introduced redundancy \cite{48Sha}.

We describe the connections (active and inactive) between loads ${\bm x}$ and firm generators ${\bm y}$ via ${\bm \sigma}=(\sigma_{i\alpha}=0,1|\{i,\alpha\}\in{\cal G})$, where ${\cal G}$ is the bi-partite graph accounting for all generators, consumers and connections.  If $\sigma_{i\alpha}=1$, the connection between load $i$ and generator $\alpha$ is active, otherwise it is inactive.  We say that a given configuration of loads ${\bm x}$ and renewable generation ${\bm y}$ is satisfiable (SAT) if there exists a matching ${\bm \sigma}$ such that the following set of conditions are simultaneously satisfied
\begin{eqnarray}
&& \forall i\in{\cal G}:\quad \sum_{\alpha\in\partial i}\sigma_{i\alpha}=1,\label{mat_cond}\\
&& \forall \alpha\in{\cal G}:\quad \sum_{i\in\partial \alpha}\sigma_{i\alpha} (x_i-z_i)\leq \hat{y}\, , \label{cap_cond}
\end{eqnarray}
where $\partial i$, respectively $\partial \alpha$, stand for all the nodes to which $i$, respectively $\alpha$, is connected.  Eq.~(\ref{mat_cond}) enforces that loads $i$ are only actively connected to a single generator, and Eq.~(\ref{cap_cond}) enforces that the sum of all loads, net local renewable generation, connected to generator $\alpha$ does not exceed the capacity $\hat{y}$ of that generator.
If the reverse is true, i.e. there exists no valid ${\bm \sigma}$ with all Eqs.~(\ref{mat_cond},\ref{cap_cond}) satisfied, we say that ${\bm x}$ is unsatisfiable (UNSAT).

Let us note here that we can also think about the current loopy power grid as the graph with $M(D+R)$ lines. Condition (\ref{mat_cond}) is in fact not necessary. However, if there are loops in the resulting network of active (switched on) links than the Kichhoff's circuit laws have to be ensured. As long as the resulting network is a tree the power flow is equivalent to general commodity flow. Requiring condition (\ref{mat_cond}) can then be regarded as restricting ourselves to the configurations of switches for which the Kichhoff's circuit laws can be ignored. It is important to notice that ignoring condition (\ref{mat_cond}) can only make the satisfiable phase larger. We require  (\ref{mat_cond}) to hold to facilitate our analysis.

First, we aim to solve the decision problem: is the given configuration ${\bm x}$ SAT or UNSAT?
Furthermore, if ${\bm x}$ is SAT, we would like to find at least one valid switching solution, ${\bm\sigma}$. As discussed below, both problems can be stated for a given grid or, alternatively,  can be considered "in average" for ensemble of networks.

\section{Preliminary Considerations}
\label{sec:results}

Let us start discussing some simple bounds on the level of firm generation $\hat{y}$ required to guarantee a SAT configuration. First of all, the total average production has to be larger than the total average  net consumption. This yields for the homogeneous distribution (\ref{Pz}) of renewables
\be
    \frac{\hat y}{D} \ge (1-\frac{\hat z}{2})\, ,  \label{cond_z}
\ee
and for the inhomogeneous distribution (\ref{Pz2})
\be
    \frac{\hat y}{D} \ge (1-f \hat z)\, .  \label{cond_f}
\ee

Another set of bounds can be obtained by considering the local structure of the network.
If $R=0$, we have a simple disconnected grid where every generator has $D$ consumers.  To guarantee that all $D$ consumer loads in each group are always SAT without any generator exceeding its capacity $\hat{y}$ and without implementing load shedding requires that
\be
      \frac{\hat{y}}{D} \ge \left(1+\frac{\Delta}{2}\right)  \ .\label{separated}
\ee
Eq.~(\ref{separated}) reflects the rare occurrence when every load within a local group of $D$ consumers is at its maximum $1+\Delta/2$ and all renewable generation at the same nodes are zero (either because all the $z_i$ are zero in the homogeneous model or because all the $f_i$ are zero in the inhomogeneous model).  Therefore, within this simplified model and in the absence of load shedding, the capacity value of the renewable generators is zero because they do not reduce the required level of firm generation relative to the case with no renewable generation, i.e. $\hat{z}=0$.

This conclusion does not imply that the renewable generators have no energy value; they can still be utilized to offset some of the firm generation $y$ within some of the local clusters of a firm generator and $D$ consumers.  However, each local cluster  must still have the maximum firm-generation capacity $\hat{y}$ given in Eq.~(\ref{separated}) to avoid {\it all} possibility of an overload.

Allowing one consumer in each cluster an ancillary connection to a remote generator ($R=1$) yields the same result.  Consider two local clusters; call them $A$ and $B$.  One local load in cluster $B$ is also connected to generator $A$.  Additionally, a local load in cluster $A$ is connected to some other remote generator.  We assume that each of the $D-1$ singly-connected consumers in $A$ and $B$ and the one shared consumer all experience the maximum load $1+\Delta/2$ and zero renewable generation.  No matter how the shared consumer and one other doubly-connected consumer in $A$ are switched, either generator $A$ or $B$ will always have $D$ connected consumers.  Therefore, the minimum allowable firm generation capacity $\hat{y}$ is the same as in Eq.~(\ref{separated}).

For a level of redundancy of $R\ge2$, the local reasoning is only a bit more complicated.  Consider again the two clusters $A$ and $B$. There are order $N$, $O(N)$, pairs of clusters in the network that share one consumer, and $O(1)$ pairs that share two consumers. The probability that two clusters share three or more consumer is vanishingly small, $o(1)$, for arbitrarily large but finite $R$. If two clusters share one or two customers, the locally best switching that can be implemented will
result in at least one of the two generators supplying at least $D-R+1$ customers.
In the rare, but finite probability, event that all the singly-connected and shared consumers within the $A$ and $B$ clusters are at the maximum load $1+\Delta/2$ and at the minimum renewable generation of zero, the minimum allowable firm generation that still yields a SAT state is
\be
      \frac{\hat{y}}{D} \ge \left(1-\frac{R-1}{D} \right)  \left(1+\frac{\Delta}{2}\right)  \ .\label{meshed}
\ee

Eq.~(\ref{meshed}) shows that, within the assumptions of our simplified model, the maximum fraction of firm generation that can be displaced by intermittent renewable generation is $(R-1)/D$, and $R$ must be greater than or equal to 2 before any capacity benefit is realized.  We note that this result is achieved solely via switching of ancillary connections.  Other methods can and should be deployed to mitigate fluctuations and allow renewable generation to claim a generation capacity factor.

Eq.~(\ref{meshed}) is obviously not exact as it only sets a bound on the maximum amount of firm generation that could be displaced by intermittent renewable generation.  The arguments resulting in Eq.~(\ref{meshed}) were based on the worst-case conditions local to a pair of firm generators, and it does not provide any hint as to how much renewable generation must be installed to reach this limit.  To answer this question precisely, we turn to an accurate computational approach described in the next Section.

\section{SAT-UNSAT transition}
\label{sec:BP}

To solve our model, defined above, we use the cavity method  and its computational realization via
population dynamics, introduced in the statistical physics of disordered systems
\cite{85MP,01MP} and recently adapted to the analysis of Shannon (phase) transitions in
constraint satisfaction \cite{02MPZ} and error-correction \cite{08RU,09MM}. This method
explores the famous fact that the Bethe-Peierls \cite{35Bet,36Pei} or Belief Propagation
\cite{63Gal,88Pea} (BP) scheme exactly solves probabilistic models on graphs without
loops (a tree). The method allows evaluation of the ensemble averages over
configurations of allowed discrete switchings on the grid that balance load with
generation.

The BP approach to a similar problem, however not accounting for a possibility of renewable generation on the consumer side, has been explained in \cite{09ZC}.  Here, we provide a brief overview of the technique.  In the asymptotic limit of an infinite system, for which $N\to \infty$ while $D$ and $R$ are $O(1)$, the interconnected grid is locally tree-like. Therefore, the BP approach for evaluating the  generalizations of condition (\ref{meshed}) is expected to be asymptotically exact (unless the system exhibits a glass transition as in \cite{01MP,02MPZ}, however, we did not observe any signs for this breakdown of the asymptotic exactness).
We introduce the following set of {\it marginal probabilities}:
$\psi_1^{\alpha\to i}$ is the probability that generator $\alpha$ is satisfied given that the edge $(i,\alpha)$, connecting $\alpha$ with his consumer-neighbor on ${\cal G}$, is in the active state, i.e. $\sigma_{i\alpha}=1$.
$\psi_0^{\alpha\to i}$ is the probability that generator $\alpha$ is satisfied given that the edge $(i,\alpha)$, where
 $(i,\alpha)\in{\cal G}_1$, is inactive, i.e. $\sigma_{i\alpha}=0$.
$\chi_1^{i\to \alpha}$ is the probability that $i$ is satisfied (i.e. it is connected to exactly one generator) given that
 the edge $(i,\alpha)$, where
 $(i,\alpha)\in{\cal G}_1$, is active, i.e. $\sigma_{i\alpha}=1$.
$\chi_0^{i\to \alpha}$ is the probability that $i$ is satisfied (i.e. it is connected to exactly one generator) given that
 the edge $(i,\alpha)$, where
 $(i,\alpha)\in{\cal G}_1$, is inactive, i.e. $\sigma_{i\alpha}=0$.
BP  relates these marginal probabilities to each other assuming that the relations are graph local, i.e. as if the graph would  contain no loops. The resulting BP equations are
\begin{align}
          \chi_1^{i\to \alpha} &= \frac{1}{Z^{i\to \alpha}} \prod_{\beta\in \partial i \setminus \alpha} \psi_0^{\beta\to i}\, , \label{BP1}\\
         \chi_0^{i\to \alpha}& = \frac{1}{Z^{i\to \alpha}} \sum_{\beta\in \partial i \setminus \alpha} \psi_1^{\beta\to i}
         \prod_{\gamma\in \partial i \setminus \alpha,\beta} \psi_0^{\gamma\to i}  \, , \label{BP2}\\
          \psi_1^{\alpha\to i}& =  \frac{1}{Z^{\alpha\to i}}  \sum_{{\bm\sigma}_{\partial\alpha\setminus i\alpha}}
          \theta(\hat y-w_{i}-\sum_{j\in \partial \alpha \setminus i} \sigma_{j\alpha} w_{j}) \prod_{j\in \partial \alpha \setminus i} \chi_{\sigma_{j\alpha}}^{j\to \alpha}\, , \label{BP3}\\
 \psi_0^{\alpha\to i}& \!=\!  \frac{1}{Z^{\alpha\to i}}  \sum_{{\bm\sigma}_{\partial\alpha\setminus i\alpha}}\!\!
  \theta(\hat y\!-\!\sum_{j\in \partial \alpha \setminus i} \sigma_{j\alpha} w_{j})\!\! \prod_{j\in \partial \alpha \setminus i} \chi_{\sigma_{j\alpha}}^{j\to \alpha} \, ,\label{BP4}
\end{align}
where $w_i=x_i+z_i$, $\partial i\setminus\alpha$ is a standard notation for the set of generator nodes linked to consumer $i$, however excluding generator $\alpha$, and similarly $\partial\alpha\setminus i$ stands for the set of consumer nodes linked to generator $\alpha$ excluding consumer $i$. The sum over ${\bm\sigma}_{\partial\alpha\setminus i\alpha}$ is over all values $\{0,1\}^{|\partial_\alpha\setminus i|}$. In Eqs.~(\ref{BP1}-\ref{BP4}), $Z^{i\to \alpha}$ and $Z^{\alpha\to i}$ are normalizations ensuring that, $\chi^{i\to \alpha}_1+\chi^{i\to \alpha}_0=1$ and
$ \psi_1^{\alpha\to i} + \psi_0^{\alpha\to i} =1$. The $\theta(\cdot)$ is the step function enforcing the generation constraints. It is unity if the argument is positive and zero otherwise. The probability for the link/edge $(i,\alpha)$ to be active, stated in terms of the related $\psi$ and $\chi$, is
\be
       p(i,\alpha)= \frac{\psi_1^{\alpha\to i}\chi_1^{i\to \alpha}}{\psi_1^{\alpha\to i}\chi_1^{i\to \alpha}+
       \psi_0^{\alpha\to i}\chi_0^{i\to \alpha}}\, ,\label{marg}
\ee
where $\sum_{\alpha \in \partial i} p(i,\alpha)=1$.

The Bethe entropy, defined as the logarithm of the number of possible SAT configurations, is
\begin{align}
   S_{\rm Bethe} &= \sum_{\alpha} \log(Z^{\alpha})  + \sum_{i} \log(Z^{i})  - \sum_{(i,\alpha)} \log(Z^{i\alpha})\, , \\
   Z^{\alpha} &=  \sum_{{\bm\sigma}_{\partial\alpha}=\{0,1\}^{|\partial_\alpha|}}
    \theta(\hat y-\sum_{i\in \partial \alpha} \sigma_{i\alpha} w_{i}) \prod_{i\in \partial \alpha} \chi_{\sigma_{i\alpha}}^{i\to \alpha}  \, , \\
 Z^{i}&=  \sum_{\alpha\in \partial i } \psi_1^{\alpha\to i}  \prod_{\beta\in \partial i \setminus \alpha} \psi_0^{\beta\to i}  \, ,  \\
 Z^{i\alpha}& =\psi_1^{\alpha\to i}\chi_1^{i\to \alpha}+\psi_0^{\alpha\to i}\chi_0^{i\to \alpha}\, .
\label{entr}
\end{align}
The entropy $S_{\rm Bethe}$ is extensive, $O(N)$, and self-averaging, i.e. the distribution of $S_{\rm Bethe}$ is concentrated around its mean value with dispersion being $o(N)$ at $N\to\infty$.

The fixed point of the BP Eqs.~(\ref{BP1}-\ref{BP4}) and the corresponding entropy (\ref{entr}) can be obtained by solving Eqs.~(\ref{BP1}-\ref{BP4}) iteratively for a given instance of the problem. Repeating these simulations many times for different instances of the graph and load ensembles, one can also calculate the average Bethe entropy. However, the average behavior, corresponding to the limits of the infinite graph, can be obtained more efficiently via the population dynamics technique. This technique offers a computationally efficient sampling from the distribution of the marginal probabilities $\chi_{\sigma}$ on infinite random graphs and subsequent evaluation of the average Bethe entropy. We will not explain here details of the Population Dynamics approach referring the interested reader to \cite{01MP,08Zde,09ZC} for further details.

Implementing the population dynamics method we observed three possible outcomes
(a) SAT phase: The Bethe entropy is positive suggesting that the number of SAT configuration (valid redistribution of the demand over the generators) is exponential in the system size.
(b) UNSAT phase, type 1: The Bethe entropy is negative, suggesting that there is almost surely no valid redistribution of demand over the generators.
(c) UNSAT phase, type 2: A contradiction is encountered in the BP equations, formally correspondent to zero values for the  normalizations in (\ref{BP1}-\ref{BP4}). We conclude that the demand is incompatible with the graph and respective generator assignment.

We shall also recall here that the SAT-UNSAT transition is actually an abrupt transition only in the asymptotic $N\to\infty$ limit. In the asymptotic limit $N\to \infty$, the SAT phase corresponds to a loss of load probability of zero and an expected unserved energy of zero.  In the UNSAT phase, the loss of load probability is one and the expected unserved energy is a finite fraction of the total demand.

\begin{figure}[!ht]
  \begin{center}
  \resizebox{0.99\linewidth}{!}{\includegraphics{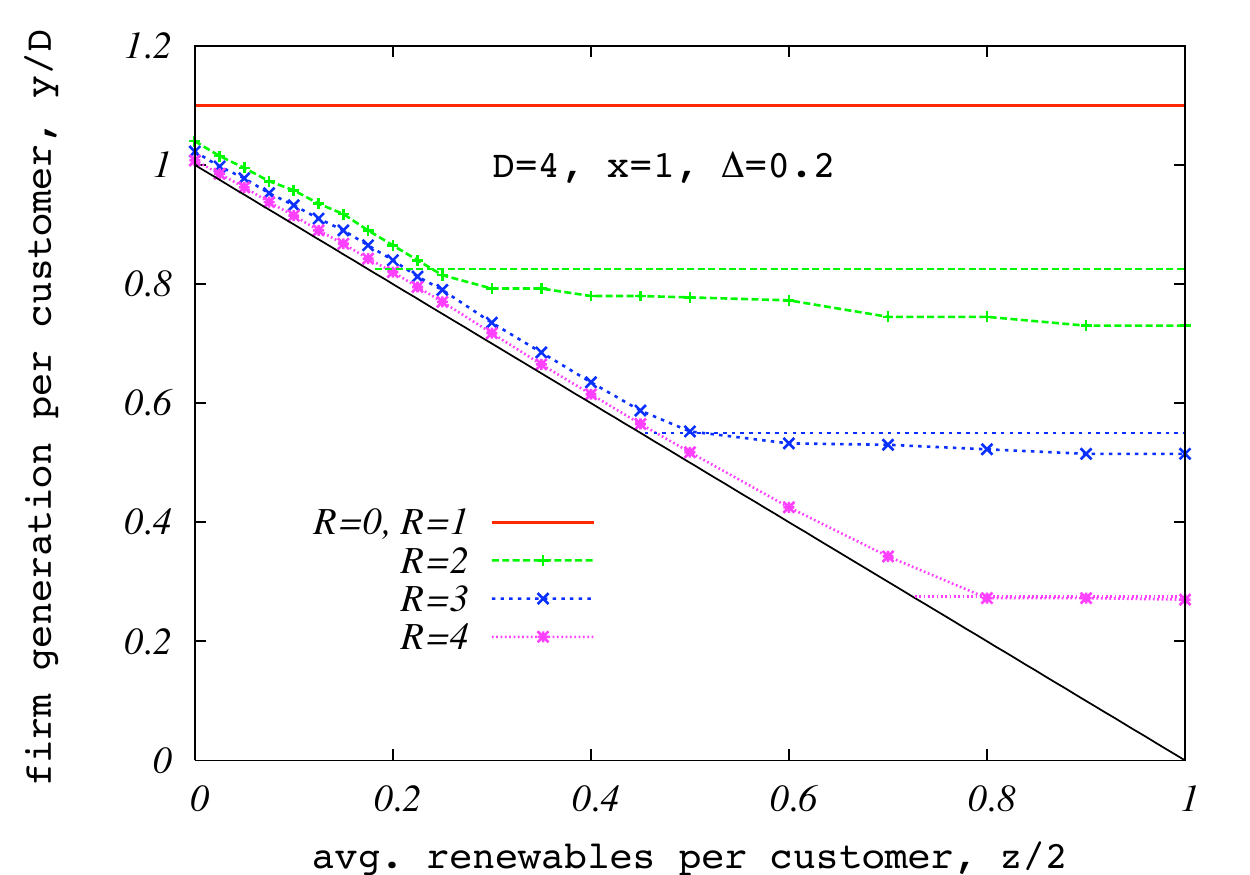}}
  \resizebox{0.99\linewidth}{!}{\includegraphics{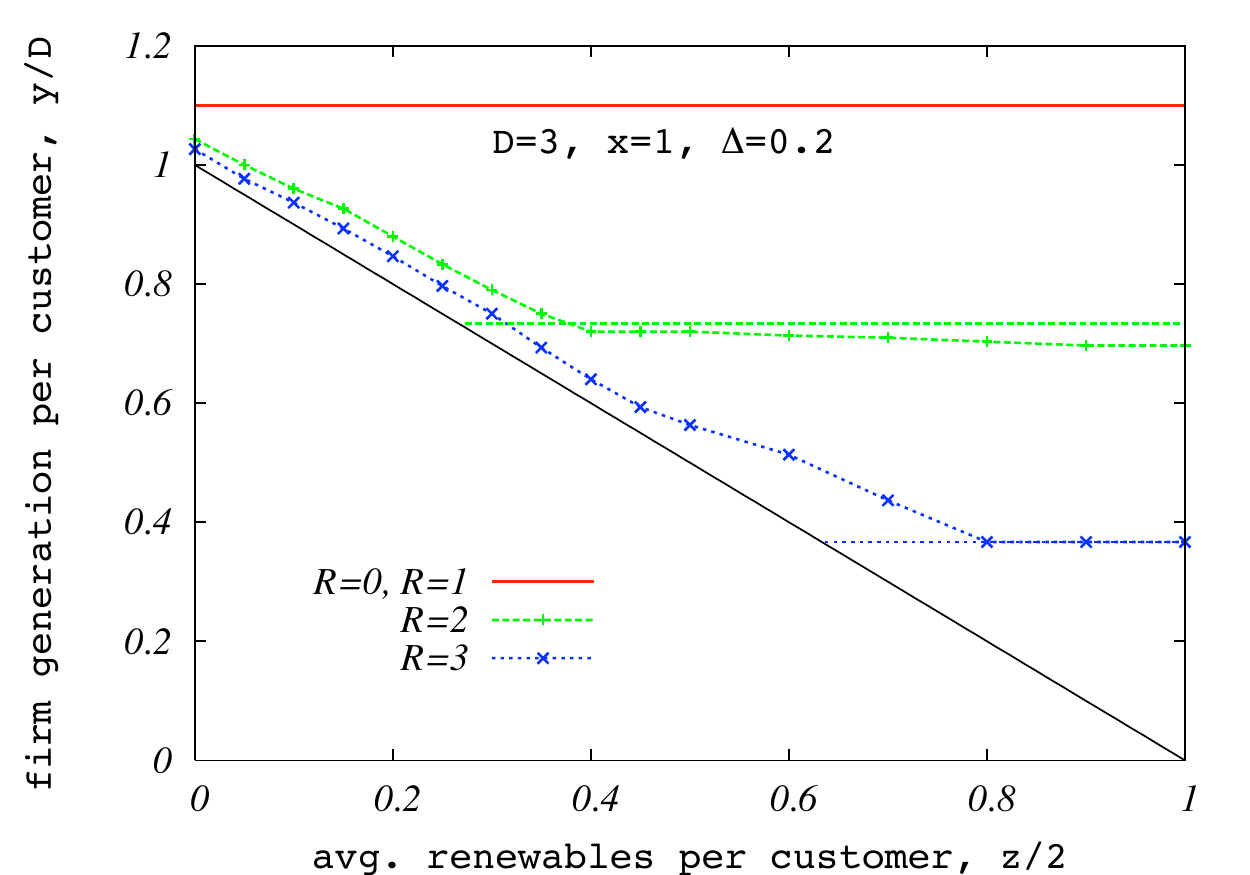}}
  \end{center}
  \caption{Results for the SAT-UNSAT threshold computed via the population dynamics for the spatially homogeneous renewable generation model at different levels of average renewable generation per consumer node . The horizontal lines correspond to condition (\ref{meshed}). The black full line corresponds to the condition (\ref{cond_z}). The colored lines connecting the data points separate the SAT and UNSAT domains lying above and below the lines, respectively. Different colors/lines/symbols correspond to different values of the redundancy parameter $R$. Top: Example for $D=4$ customers per firm generator. Bottom: $D=3$. \label{fig:hom}}
\end{figure}

\begin{figure}[!ht]
  \begin{center}
  \resizebox{0.99\linewidth}{!}{\includegraphics{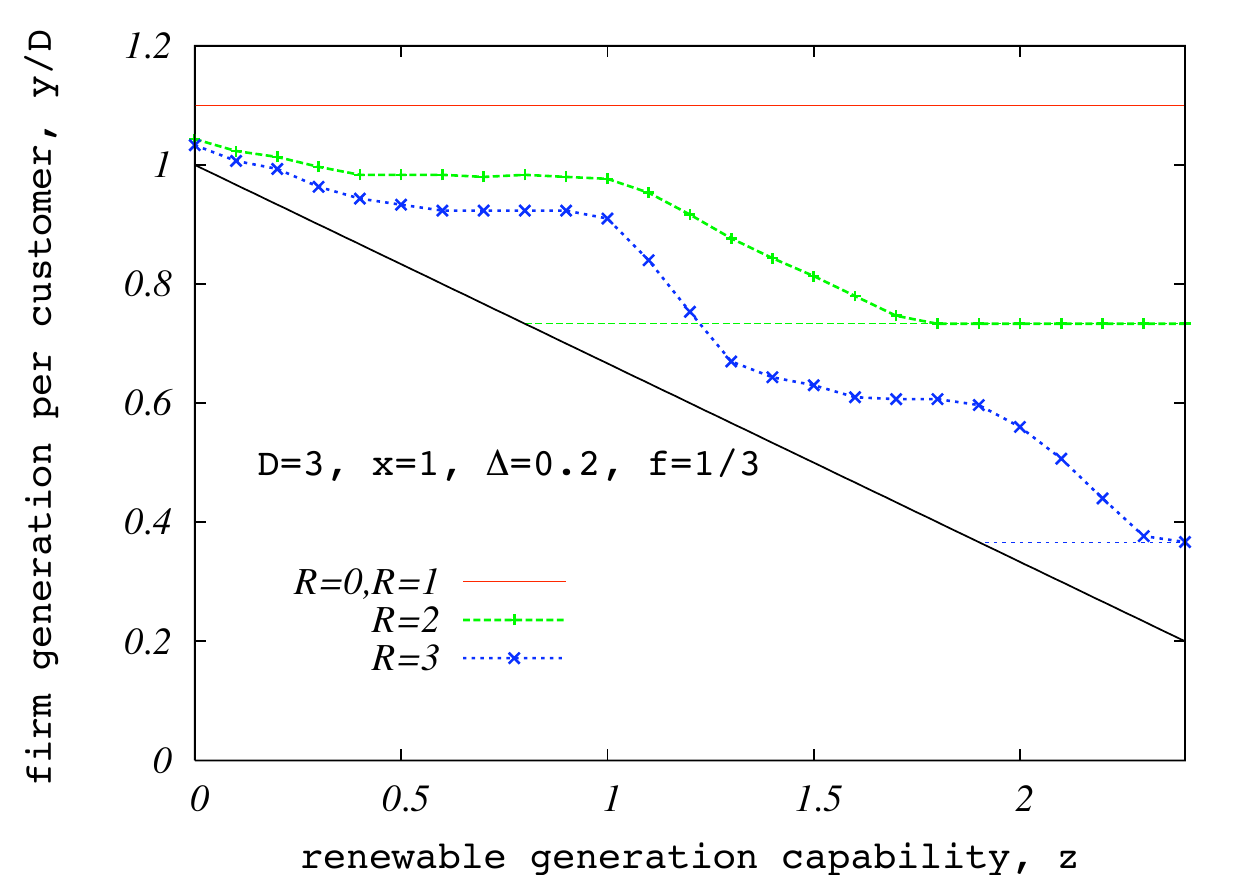}}
  \end{center}
  \caption{The necessary firm generation per customer for the inhomogeneous model with $f=1/3$ of customers producing renewables for different levels of renewable generation capability (i.e. nameplate capacity). The full black line depicts condition (\ref{cond_f}) and the horizontal lines condition (\ref{meshed}) for different values of redundancy $R$. The colored lines connecting data points separate the SAT and UNSAT domains lying above and below the lines, respectively. Different colors/lines/symbols correspond to different values of redundancy $R$. \label{fig:inhom1}}
\end{figure}

\begin{figure}[!ht]
  \begin{center}
  \resizebox{0.99\linewidth}{!}{\includegraphics{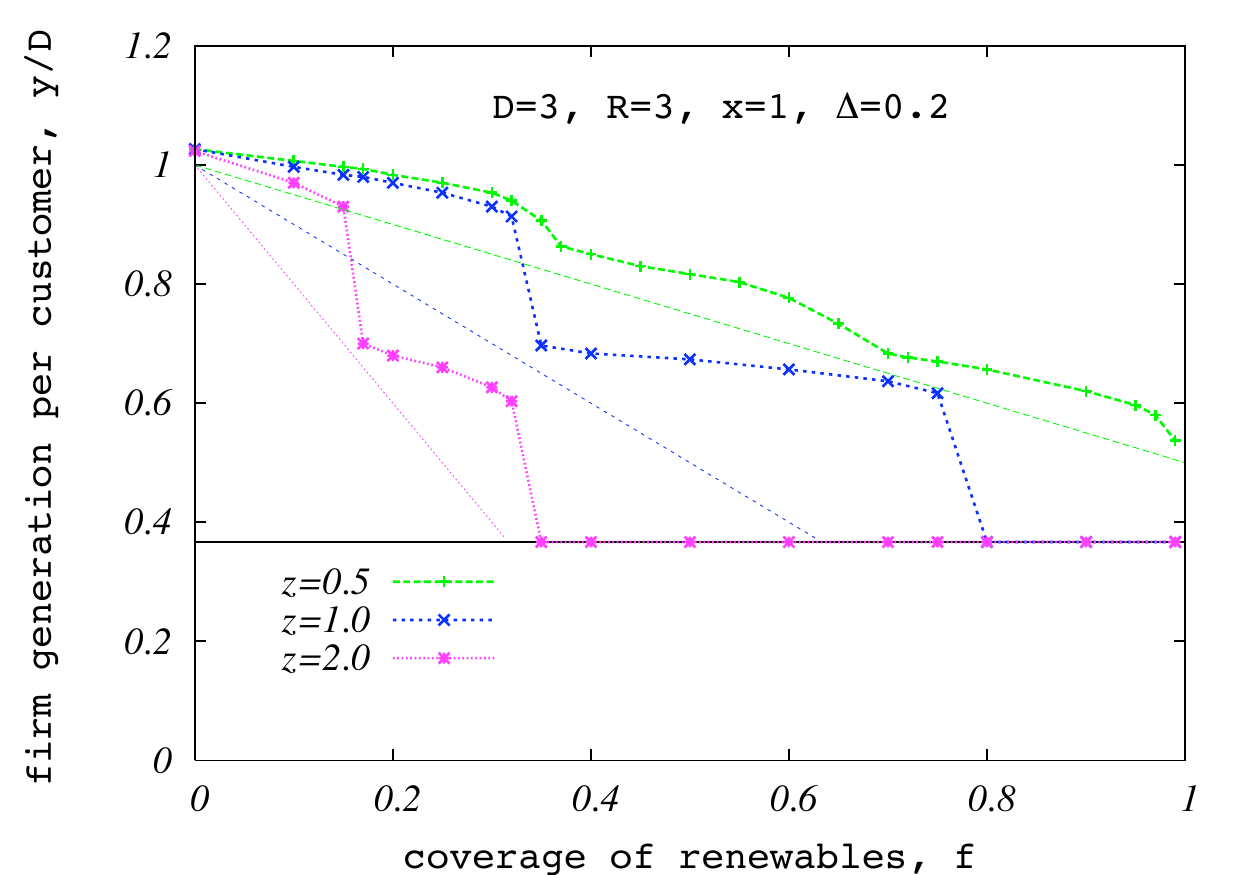}}
  \end{center}
  \caption{The necessary firm generation per customer for the inhomogeneous model with fixed installed renewable generation capability $\hat z=0.5, 1.0, 2.0$ as a function of the fractional coverage $f$. The full black horizontal line this time depicts condition (\ref{meshed}) and the colored lines with slopes $\hat z$ represents (\ref{cond_f}). The colored lines, connecting the population dynamics data points, separate SAT and UNSAT domains which lay above and below of the lines, respectively. \label{fig:inhom2}}
\end{figure}

Figures~\ref{fig:hom}, \ref{fig:inhom1} and  \ref{fig:inhom2} show results of the population dynamics simulations for the homogeneous (wind farm) and
inhomogeneous (solar) models.  Figure~\ref{fig:hom} depicts the required firm generation per customer, $\hat y/D$, as a function of average renewable generation $\hat z/2$ per customer in the homogeneous model (\ref{Pz}) for different levels of redundancy $R$.  The black full line corresponds to the condition (\ref{cond_z}). The horizontal lines to condition (\ref{meshed}) for different values of redundancy $R$. The lines connecting the symbols indicate the boundaries of the respective SAT and UNSAT phases obtained via the population dynamics. These boundaries are established by traversing phase space by decreasing $\hat{y}$ and catching the value where the UNSAT conditions are first observed. For larger values of $\hat z$ the data points are slightly below the boundary line (\ref{meshed}) due to numerical imprecision (i.e. larger population sizes are needed to encounter the rare case responsible for condition (\ref{meshed})).

In the homogeneous model and for the parameter values used in Fig.~\ref{fig:hom}, a fraction close to one of the average renewable generation can be utilized as capacity for level of redundancy $R\ge 2$. Condition (\ref{meshed}) combined with (\ref{cond_z}) gives a rough value of average renewable generation
\begin{equation}
\frac{\hat{z}}{2}=\left (\frac{R-1}{D}\right ) - \frac{\Delta}{2}\left (1- \frac{R-1}{D} \right )
\end{equation}
above which adding more renewable generation into the grid does not add any extra capacity value. This threshold increases with the level of redundancy.

Figure.~\ref{fig:inhom1} shows the necessary firm generation per customer for the inhomogeneous model (\ref{Pz2}) with $f=1/3$ (a third of customers with renewable generation) as a function of the amount of the renewable generation capability $\hat z$. Again the full black line depicts the global condition (\ref{cond_f}) and the horizontal lines the local condition (\ref{meshed}) for different values of redundancy $R$. The lines connecting the data points are drawn from the population dynamics solution for BP equations. The drop-plateau structure of the SAT-UNSAT threshold is related to the $\hat z$-dependent number of customers which can be served by one producer. Qualitatively, the change in behavior seen around $\hat z \approx 1$ is related to customers with renewables becoming independent of the producers, and another visible change at $\hat z \approx 2$ can be explained by the fact that consumers with renewables start to supply power reliably to other consumers which do not have renewable capability.

Fig.~\ref{fig:inhom2} shows the necessary firm generation per customer computed via the population dynamics technique for the inhomogeneous models (\ref{Pz2}). In this Figure, the renewable production capability $\hat z$ is fixed, and the fraction of consumers $f$ with renewable capability is variable.  The three curves are for different values of $\hat z$. One observes that the larger the renewable production $\hat z$, the more one is able to displace firm generation. On the other hand, for smaller values of $\hat z$ the fraction of production lost in order to support fluctuations is smaller. The drop-plateau structure of the curves can be interpreted.
The drop at about $f=1/3$ for $\hat z=1.0$ and $\hat z=2.0$ is related to the fact that when more than a third of customers has a renewable production then on average one customer per producer ($D=3$) can become independent (for $\hat z =1$) or even supply another customer (for $\hat z =2$). Similar reason applies for the drop at about $f=1/6$ for $z=2.0$, where on average one customer per two producers can supply another customer.

To conclude, this study of the average Bethe entropy shows that networks with added ancillary lines are able to withstand fluctuations in the renewable generation turning this generation into capacity.  This effect is amplified with increasing $R$. The structure of the satisfiable region depends on the distribution of demand and renewables production. The drop-plateau structure of the curves in the inhomogeneous model suggests that even in more realistic systems one might expect a phase where adding a few more renewable resources does not contribute much additional capacity and a phase where adding a few more renewables can, on contrary, add more capacity than the amount of additional renewable generation.  (when the slope of the SAT/UNSAT curve is steeper than the slope of lines (\ref{cond_f})). We did not observe this structure in the homogeneous model (Figs.~\ref{fig:hom}).  The results of our simplified models are guides to the type of generic behavior expected in more realistic models which should be investigated in detail to determine accurate phase diagrams.

\section{Control Algorithm} \label{sec:Control}
Belief Propagation provides an efficient message-passing control algorithm for assigning loads to generators.  In the spirit of \cite{02MPZ}, we illustrate that the BP-based decimation algorithm is able to find valid configurations within the SAT domain.  The decimation algorithm works as follows:\\
{\bf DECIMATION:}\\
1 {\bf repeat} (\ref{BP1}-\ref{BP4}) $n$ times\\
2 compute the marginals (\ref{marg})\\
3 select the most biased edge and fix its state\\
4 cut the fixed edge from the graph\\
5 {\bf until} solution or contradiction is found.

The BP-based decimation algorithm updates Eqs.~(\ref{BP1}-\ref{BP4}) iteratively by passing messages from generators to consumers and back. After fixed number of steps, the most biased consumer is chosen, the more probable value for its consumption is assigned, the graph is reduced, and the procedure is repeated. Note that updating Eqs.~(\ref{BP1}-\ref{BP2}) takes $2^{2R}$ steps per message thus making the algorithm exponential in $R$. However, building new connections is expensive and $R$ will not be very large in realistic cases. The algorithm is quadratic in the number of consumers, but linear computational time is achievable if a finite fraction of customers is decimated at each step, just as in \cite{02MPZ}. \\

\begin{figure}[!ht]
\begin{center}
\resizebox{0.99\linewidth}{!}{\includegraphics{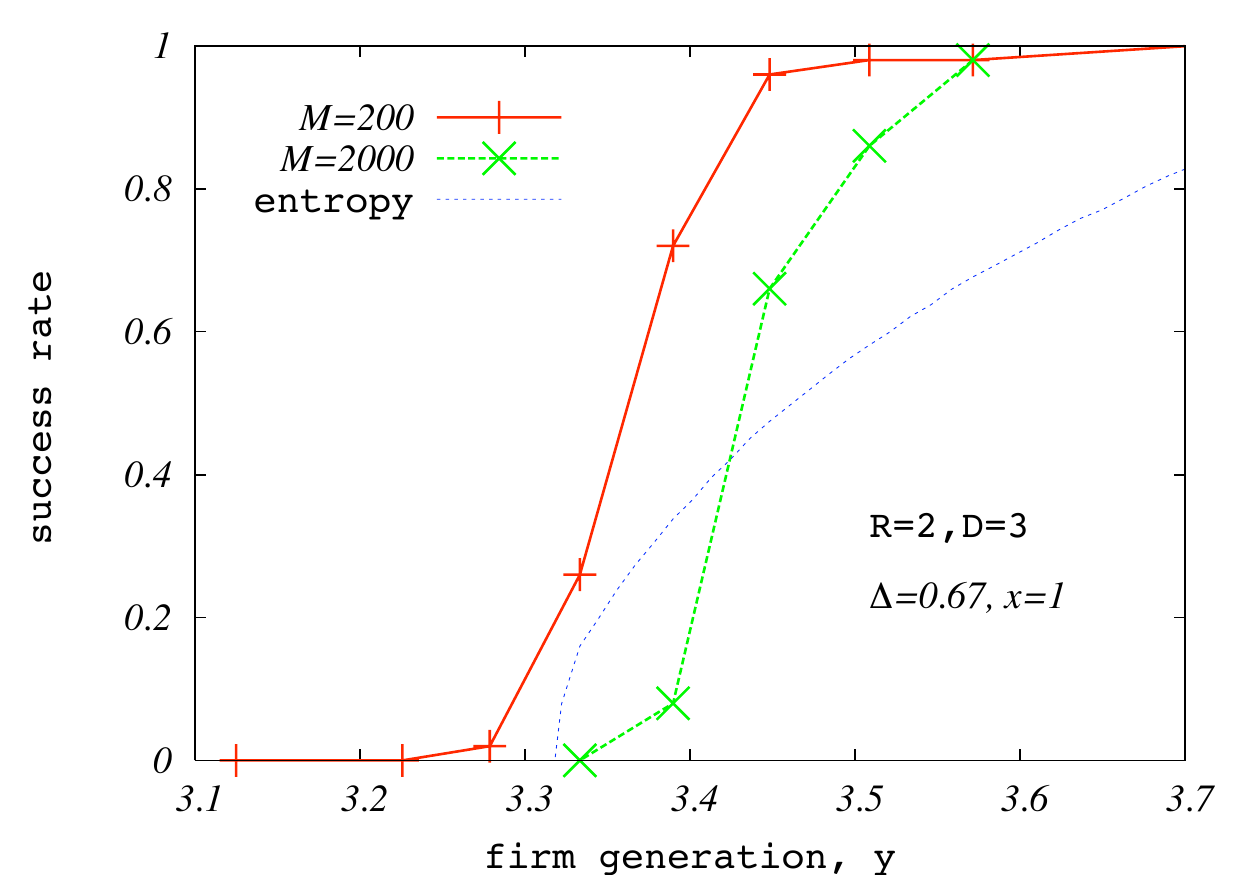}}
\end{center}   \caption{ \label{fig_dec} Performance of the BP-based decimation
algorithm. The data are for networks with $M$ firm generators, $n=5$ (five iteration per a cycle of the decimation procedure) and $R=2$, $D=3$, $\Delta=0.67$. An average over $50$ random instances is taken, and the fraction of successful runs is plotted against the production cap $\hat y$. The Bethe entropy-based (asymptotic) curve is drawn dashed for comparison (the actual value for the curve is not related to the success rate), suggesting that in the limit of $N\to \infty$ valid configurations exist starting from $\hat y \approx 3.33$. Note that in the separated case (of $R=0$), valid configurations exist only above $\hat y \ge D(1 + \Delta/2) = 4$.}
\end{figure}

The algorithm performance is illustrated in Fig.~\ref{fig_dec} where the percentage of success in the BP-based decimation is shown. These data  average over $50$ random instances from the $R=2$, $D=3$ ensemble with a width in consumption of $\Delta=0.67$, no renewable generation, and a variable generation cap. For such a set of parameters, the separated network would require a firm generation capacity of
$\hat y \ge 4$, whereas with two consumers per producer connected to two producers the minimum firm generation decreases to $\hat y \approx 3.33$.

\section{Summary and Path Forward}
\label{sec:Con}

This manuscript expands upon \cite{09ZC} describing a Belief-Propagation approach to determining the satisfiability of load and generation balance in the absence of load shedding for electrical grids including redundant ancillary connections.  Here, we have included intermittent renewable generation at consumer nodes and computed the minimum level of firm generation required to avoid an overload of any the firm generators.  The amount of firm generation that can be displaced by the combination of renewable generation and intelligent switching of the ancillary lines equates to the generation capacity value of the renewable generators.  Our results provide general guidance about the behavior of renewable generation capacity for various levels of redundancy and renewable penetration, and we find that the addition of ancillary lines greatly increases the capacity value of the renewable generators.

The general approach we pursued in this study is based on recent developments in the field of graphical models that merge statistical physics,  computer science, optimization theory and information theory \cite{08RU,09MM}.  The approach is asymptotically exact on infinite sparse graphs and provides an efficient heuristic for graphical models on finite sparse graphs. The BP-approach is general and of a dual use: on the one hand BP is a good for asymptotic (capacity/phase-transition style) analysis, and it also generates efficient algorithmic solutions for optimization and control of the power grid.

Note that many generalizations of our analysis are very straightforward and can be used directly within the framework presented here. This includes implementing different probability distributions for demands and renewable generation (as long as the distribution support is bounded) or consideration of non-uniformity for both generators and consumers, i.e. varying production caps for either firm or renewable generators and introducing distinct distributions of demands for different consumers.  Also, our simplified network can be easily extended in many ways. Our equations are straightforwardly valid for every locally tree-like random network. Another case, which allows very natural and straightforward generalization for all the statements made in this manuscript, corresponds to breaking the equivalence between different edges in the graph and thus assigning nonuniform weights to them.  These weights may e.g. represent cost of construction, geographical length, proxy for losses, cost of exploitation, etc.  There are many other more realistic extensions of our model associated with the description, optimization and control of power grids which can  benefit from utilizing a graphical model approach of the kind discussed in this paper,  even though actual analysis may prove to be more involved.

Let us emphasize that the BP-based control algorithm developed in the manuscript represents a generic approach that can be used for efficient switching on any power grid, structured or not.
This switching algorithm is a heuristic which does not rely on any asymptotic assumption (which is only needed for theoretical analysis of the SAT-UNSAT transition), and as such it offers a very powerful practical control tool.

\section{Acknowledgments}
We thank members of LANL study group on "Optimization and Control Theory for Smart Grids" for many discussions. The work at LANL was carried out under the auspices of the National Nuclear Security Administration of the U.S. Department of Energy at Los Alamos National Laboratory under Contract No. DE-AC52-06NA25396.

{\small
\bibliographystyle{unsrt}
\bibliography{GridRed}
}

\end{document}